\documentclass{emulateapj}
\usepackage{graphicx}
\usepackage{subfigure}
\usepackage{longtable}

\slugcomment{Submitted to ApJ on November 2}

\begin{document}
\title{A Technique for Detecting Starlight Scattered from Transiting Extrasolar Planets with Application to HD 209458b}

\shorttitle{DETECTING SCATTERED LIGHT FROM TRANSITING EXOPLANETS}
\shortauthors{LIU ET AL.}

\author{\sc Xin Liu, Edwin L. Turner}
\affil{Princeton University Observatory, Princeton, NJ 08544, USA}

\author{\sc Norio Narita, Yasushi Suto}
\affil{Department of Physics, University of Tokyo, Tokyo 113-0033, Japan}

\author{\sc Joshua N. Winn}
\affil{Department of Physics, and Kavli Institute for Astrophysics and
Space Research, Massachusetts Institute of Technology, Cambridge, MA 02139, USA}

\author{\sc Toru Yamada}
\affil{Astronomical Institute, Tohoku University, Sendai 980-8578, Japan}

\author{\sc Bun'ei Sato}
\affil{Tokyo Institute of Technology, 2-12-1-S6-6, Ookayama, Meguro-ku, Tokyo 152-8550, Japan}

\author{\sc Wako Aoki, Motohide Tamura}
\affil{National Astronomical Observatory of Japan, 2-21-1 Osawa, Tokyo 181-8588, Japan}

\begin{abstract}
We present a new technique for detecting scattered starlight from
transiting, close-orbiting extrasolar giant planets (CEGPs) that
has the virtues of simplicity, robustness, linearity, and
model-independence. Given a series of stellar spectra obtained
over various phases of the planetary orbit, the goal is to measure
the strength of the component scattered by the planet relative to
the component coming directly from the star. We use two
complementary strategies, both of which rely on the predictable
Doppler shifts of both components and on combining the results
from many spectral lines and many exposures. In the first
strategy, we identify segments of the stellar spectrum that are
free of direct absorption lines and add them after
Doppler-shifting into the planetary frame. In the second strategy,
we compare the distribution of equivalent-width ratios of the
scattered and direct components. Both strategies are calibrated
with a ``null test'' in which scrambled Doppler shifts are applied
to the spectral segments. As an illustrative test case, we apply
our technique to spectra of HD 209458 taken when the planet was
near opposition (with orbital phases ranging from 11 to
34$\arcdeg$, where 0$\arcdeg$ is at opposition), finding that the
planet-to-star flux ratio is $(1.4\pm 2.9)\times10^{-4}$ in the
wavelength range 554$-$681~nm. This corresponds to a geometric
albedo of $0.8\pm1.6$, assuming the phase function of a Lambert
sphere. Although the result is not statistically significant, the
achieved sensitivity and relatively small volume of data upon
which it is based are very encouraging for future ground-based
spectroscopic studies of scattered light from transiting CEGP
systems.

\end{abstract}


\section{Introduction}

The discovery of the first exoplanet orbiting an ordinary main
sequence star, 51 Peg, a dozen years ago \citep{mayor95} also
revealed the first example of a largely unexpected but common
class of astronomical objects, namely close-orbiting extrasolar
giant planets (CEGPs).  The discovery of the first transiting
exoplanet system, HD 209458b also a CEGP
\citep{henry00,mazeh00,charbonneau00,brown01}, a few years later
allowed study of these enigmatic objects to move beyond orbital
parameters and minimum masses.  Although radial velocity, transit
and microlensing techniques have made the study of exoplanets one
of the most active and exciting areas of astrophysics in recent
years and have produced a wealth of additional major discoveries
and important theoretical problems \citep{perryman00,chauvin07},
the effort to understand CEGPs remains central to the field.

The light of the primary star scattered from CEGPs reflects the
nature (composition, clouds, energy balance, structure, motions
{\it et cetera}) of their atmospheres
\citep[e.g.,][]{seager98,marley99,seager00,sudarsky00,sudarsky03,burrows05};
its direct detection would thus test and discriminate among
emerging models of CEGPs.

There have been previous attempts to detect scattered starlight
from CEGPs against the much brighter direct light from the primary
via high-resolution spectroscopic searches for the scattered, and
hence Doppler-shifted, stellar absorption lines \citep[][and
references therein]{charbonneau99,leigh03}. The problem is very
difficult because the expected contribution of the scattered light
to the total light is of order $10^{-4}$ or less for known CEGPs.
Due to both the very small amplitude of the expected signal and
the unknown orbital inclination angle $i$ for non-transiting
systems, the methods so far employed rely on model-dependent
assumptions to estimate prior distributions for planetary radii,
orbital inclinations and the form of the phase function
\citep{leigh03}.  The advantage of such an approach is that it can
be applied to the brightest (but non-transiting) CEGP systems that
thus have the least photon noise. In other words and as is usually
the case, relying on a prior model allows one to maximize purely
statistical signal-to-noise (S/N) but at the price of increased
systematic uncertainty. Two previously published methods
\citep[e.g.][]{charbonneau99,cameron99} depend on a $\chi^2$
minimization applied to the whole stellar spectrum, which is not
an optimally sensitive method for this problem because it dilutes
the detectability of the scattered lines (which carry the signal
of interest) by considering the quality of the fit to the
continuum regions.

The simplest example of the advantage of studying transiting CEGP
systems is that measurement of the scattered light fraction allows
direct determination of the albedo in isolation, and not merely in
combination with the planet's radius squared (which must be
estimated from a theoretical model for non-transiting systems).
The major advantage of studying non-transiting systems is that
some of them are nearly two orders of magnitude brighter than the
brightest transiting system, thus yielding far smaller photon
statistics noise.

Here we propose a largely model-independent technique to detect
the scattered stellar absorption lines in transiting CEGP systems.
In this method, no assumptions concerning the form of the phase
function, the planet's radius, the orbital inclination {\it et
cetera} are needed to measure the planet-to-star flux ratio and
albedo.  Rather, only physical parameters of the system measured
via transit observations are required. Moreover, the technique
maximizes sensitivity by concentrating only on those portions of
the spectrum in which scattered stellar lines are known to be
present (at some amplitude to be measured).  As a practical
``worked example'' we apply this technique to {\it Subaru} HDS
spectra of HD 209458.
The required input physical parameters of the system and their
relation to the expected scattered light signal are briefly
reviewed in \S 2. In \S 3, we describe the technique in detail,
and apply it to HD 209458b, using high precision optical spectra
acquired on UT 2002 October 27, with the Subaru High Dispersion
Spectrograph \citep[HDS;][]{noguchi02}.  Results are summarized in
\S 4. Prospects for future studies are then examined based on these
results.

\section{Input Parameters}

Our technique utilizes known physical properties of the transiting
CEGP system based on prior radial velocity and transit studies.

The planet's velocity relative to the star at orbital phase $\phi$
is given by \citep[see, e.g.][]{cameron02}
\begin{equation}
V_P(\phi) = K_P \cdot {\rm sin}2\pi\phi,
\end{equation}
where $\phi = (t - T_0)/P_{orb}$ is the orbital phase at time $t$,
$T_0$ is the transit epoch, and $P_{orb}$ is the orbital period.
Its apparent radial-velocity amplitude $K_P$ about the mass center
of the system is
\begin{equation}
K_P = \frac{2\pi a}{P_{orb}}\frac{{\rm sin} i}{1 + q},
\end{equation}
where $a$ is the orbital distance, $i$ is the inclination and $q
\equiv M_P/M_{\star}$ is the mass ratio. The values of $a$ and $i$
can be measured from radial-velocity and transit-photometry data,
respectively.

If the planet is tidally locked, as expected for nearly all CEGPs,
there will be no rotational broadening. Then the scattered spectra
will be the Doppler-shifted stellar spectra, scaled by a factor of
the planet-to-star flux ratio $\epsilon$, which is usually
decomposed as \citep{charbonneau99,cameron99}
\begin{equation}\label{equ:ratio}
\epsilon(\alpha, \lambda) = p(\lambda)g(\alpha, \lambda)\frac{R_P^2}{a^2},
\end{equation}
where $p(\lambda)$ is the geometric albedo, and $g(\alpha,
\lambda)$ is the phase function, with $\alpha$ being the phase
angle which varies as ${\rm cos}\alpha = - {\rm sin}i \,{\rm
cos}2\pi\phi$. The value of $R_P$ is also measured from
transit-photometry data.

\section{Method and Application to HD 209458b}

\subsection{Overview}

The basic idea behind the method is that for any absorption line
in the primary star's spectrum we can precisely predict the
wavelength of each line in the light scattered from the planet in
each exposure; it is simply offset from the position of the
stellar line by the Doppler shift due to the planet's orbital
motion relative to the primary, as projected onto the line of
sight to the observer at the time of that exposure.  The strength
of the scattered line relative to the line in the primary's
spectrum is just the fractional contribution of the planet's
scattered light to the total light at that wavelength; in
principle,  this line strength gives a measure of the planet's
albedo at that wavelength and phase angle.

Of course, in practice, for known CEGP systems this fraction is
expected to be of order $10^{-4}$ to $10^{-5}$ and any individual
line is thus completely undetectable even in very high S/N
spectra. Our technique is then simply to superimpose and combine
pieces of the spectra at the expected wavelengths of the scattered
lines over many lines and exposures to achieve sufficient S/N to
allow a statistically significant detection. This can be done
either 1) by simply summing small fragments of the spectra
centered on these predicted locations or 2) by measuring the ratio
of scattered absorption line equivalent-widths to the
corresponding equivalent widths in the primary and then combining
the ratios over lines and exposures. We employ both variations.

In an attempt to detect tiny fractional signals against a noisy
background, it is obviously crucial to know the background rather
precisely and to characterize its noise properties accurately.  We
do both by an ``internal'' procedure using the same spectra and
same wavelength regions in which we search for the signal.
Specifically, in addition to combining the spectral fragments
defined by the expected wavelength positions of the scattered
lines, we also carry out the same analysis but with the Doppler
shift assigned to each exposure ``shuffled'' randomly among the
exposures.  In this way we can produce a large number of {\it
reference spectra} made up of the same spectral fragments taken
from the same set of exposures but each lacking any scattered line
contribution.  By combining this ensemble of reference spectra we
obtain an excellent estimator of both the background against which
we must distinguish the scattered light signal and of all of its
noise properties (including, for example, the collective effects
of weak stellar lines that cannot be detected individually).

Finally, in order to calibrate and test the method, it is very
useful to have mock or simulated data containing an artificially
injected, and thus precisely known, signal and having realistic
noise properties.  This we also manufacture internally.  The
individual spectra of the system are shifted to the planet's rest
frame at each exposure, and then multiplied by some assumed
planet-to-star flux ratio $\epsilon$ to produce a set of
artificial planet scattered light spectra. These artificial
spectra are then added back into the individual exposures. For
example, if the original spectra contain some scattered light
signal with a planet-to-star ratio $\epsilon_0$, then these
artificially injected spectra should have scattered light with a
planet-to-star ratio $\epsilon_1 = (\epsilon_0 + \epsilon)$. We
can analyze the artificially injected spectra to obtain
$\epsilon_1$ and then subtract $\epsilon_0$ from it to recover
$\epsilon$.  Successful, or failed, recovery of this known
artificial signal by application of the same technique provides a
strong test of the reliability and sensitivity of the method.

\subsection{Data Preprocessing}

As an illustrative case study, we analyze 33 Subaru HDS spectra of
the system HD 209458, taken on UT 2002 October 27, when the planet
was just out of the secondary eclipse, with the orbital phase
$11.0^{\circ} \lesssim \alpha \lesssim 33.9^{\circ}$. A detailed
description of the observations and data reduction to obtain
one-dimensional spectra can be found in \citet{winn04} and
\citet{narita05}. The wavelength coverage of the spectra analyzed
is from 554 nm to 681 nm, with a spectral resolution of $R \approx
45000$. The typical exposure time was 500 sec, and the S/N per
pixel was $\sim 350$.

The reasons for adopting exposure phases in the vicinity of the
secondary eclipse, but not extremely close to it, are two fold.
First, the strong back scattering of aerosol clouds or grains, if
any, could dramatically brighten the planet. Second, we should not
observe too close to either the primary transit or the secondary
transit, because during those events the planet's motion is nearly
in the plane of the sky and its relative radial velocity variation
is small, causing the scattered absorption lines to be blended
with the direct absorption lines.  For the HD 209458 data analyzed
here, the velocity offsets between the stellar lines and their
scattered counterparts is in the range of $27 - 80$ km s$^{-1}$,
sufficient to avoid blending in all cases.

The one-dimensional spectra are first normalized. We
fit an 11$^{th}$-order-{\sf spline3} function to obtain the blaze
function of each order separately, and then merge the blaze
functions of all the orders. The un-normalized spectra of each
order are also merged in the same way and divided by the merged
blaze function. Spectra are re-sampled and dispersion-corrected to
logarithmically spaced bins after merger. All exposures are
normalized separately, as their blaze functions vary due to
instrumental variations \citep{winn04,narita05}. We note that the
method to be discussed in Section \ref{sec:strat1} is also
applicable to the un-normalized spectra. The normalization however
simplifies the EW test to be described in Section
\ref{sec:strat2}. The reason we use an 11$^{th}$-order-{\sf
spline3} function for the fitting is to remove the large-scale
profiles of the spectra without diluting the weak lines containing
the scattered light signal.

We adopt the physical parameters of the system HD 209458 from
\citet{knutson07}, as summarized in Table \ref{tab:para}, along
with the Modified Julian Day (MJD) at the central time of each
exposure, to predict $V_P(\phi)$.

\begin{deluxetable}{lc}
\tablewidth{0pt} \tabletypesize{\footnotesize}
\tablecaption{Physical parameters of the system HD209458 from
\citet{knutson07}.\label{tab:para}}
\tablehead{
\colhead{Parameter} &
\colhead{Value and references}
}
\startdata
    Mass of the star: $M_{\star}$ ($M_{\odot}$)   &    1.101$^{+0.066}_{-0.062}$    \\
    Radius of the star: $R_{\star}$ ($R_{\odot}$)   &    1.125$^{+0.020}_{-0.023}$ \\
    Mass of the planet: $M_P$ ($M_J$)   &    0.64$\pm$0.06\\
    Radius of the planet: $R_P$ ($R_J$)   &   1.320$^{+0.024}_{-0.025}$ \\
    Orbital inclination: $i$ (degrees)   &    86.929$^{+0.009}_{-0.010}$ \\
    Orbital period: $P_{orb}$ (days)   &    3.52474859$\pm$0.00000038 \\
    Transit epoch: $T_0$ (HJD)   &    2,452,826.628521$\pm$0.000087 \\
\enddata
\end{deluxetable}

\subsection{Strategy I: Summing Scattered Light Lines}\label{sec:strat1}

In this section, we search for the scattered light by aligning and
combining the imprinted signals over a large set of stellar
absorption lines as well as over exposures. The scattered signals
are aligned by Doppler-shift correcting each exposure according to
the corresponding $V_P(\phi)$. We also make a set of reference spectra
by combining the same set of exposures, which, on the other hand,
are Doppler-shift corrected with the $V_P(\phi)$ of another
exposure. Therefore, extra absorptions around the predicted
scattered-light centers would indicate a detection.

\subsubsection{Details of the Strategy-I Procedure}

\begin{figure}
\epsscale{1.1}
\plotone{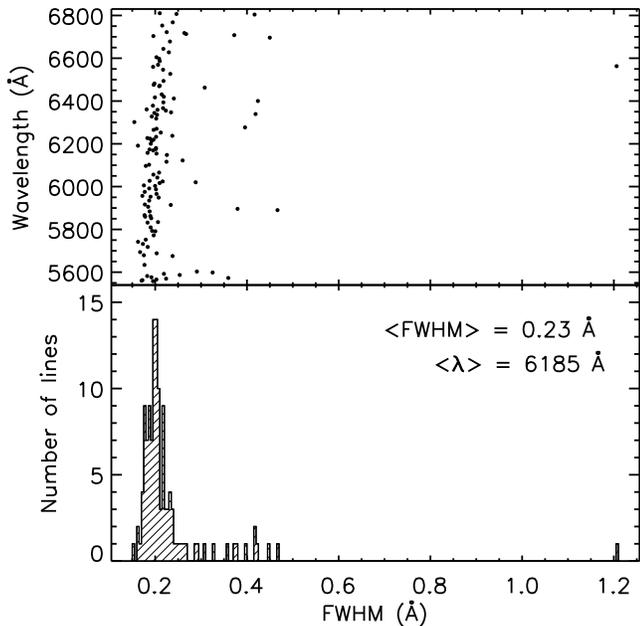} \caption{The 123 selected stellar absorption lines.
Wavelength versus FWHM ({\it top}),
and the distribution of FWHM ({\it bottom}) .\label{fig:fwhm}}
\end{figure}

First, 33 pre-processed spectra $S^0_{{\rm log}\lambda}$ are
Doppler-shift corrected according to their predicted corresponding
radial velocities. We use a logarithmic wavelength scale, and the
super script ``0'' to refer to the star's rest frame and ``Z'' for
the planet's. Then the shifted spectra, $S^Z_{{\rm log}\lambda}$,
are corrected to have the same dispersion as that of $S^0_{{\rm
log}\lambda}$. Here and throughout this study, spectra are always
corrected to this standard dispersion after being shifted.

Second, a set of suitable stellar absorption lines were selected
according to the following criteria:  1 - The line's full-width at
half-maximum (FWHM) must be larger than the threshold value
FWHM$_{th}=0.15 {\rm \AA}$, a value chosen by experimentation to
maximize sensitivity.  (The final result is, however, rather
insensitive to the exact value of FWHM$_{th}$.) 2 - The line's
blue\footnote{Due to the planet's phases, its scattered spectra in
all the 33 exposures we used in this study would be blue-shifted
relative to the stellar spectra.}-ends extending to a wavelength
range, which corresponds to the largest radial velocity among all
the exposures, needs to be relatively free of strong stellar
lines. This is required to avoid contamination of any detection by
strong stellar lines. The wavelengths of selected lines and the
distribution of their FWHMs are displayed in Figure
\ref{fig:fwhm}.

Third, small fragments ($\Delta \lambda \approx$ 1.7 ${\rm \AA}$
each\footnote{$\Delta \lambda \approx$ 1.7 ${\rm \AA}$ was chosen
to be big enough to cover several typical stellar line widths.})
of the spectra are averaged together, centering on the predicted
scattered features and over all the selected lines and exposures.
The combined normalized counts as a function of wavelength
difference (from the predicted line centers) can be written as
\begin{equation}
C^Z(k) = \frac{1}{N_{e}N_{l}}\sum_{e = 1}^{N_{e}} \sum_{l =
1}^{N_{l}} S^{Z_e}[{\rm log}(\lambda_{k, el}) - {\rm log}(\lambda_{0, el})],
\end{equation}
where $k = -50, -49, ..., 49, 50$, and represents the $k$-th pixel
on the bluer (redder) side of the scattered-light center ($k =
0$), if $k$ is negative (positive); ${\rm log}(\lambda_{k + 1,
el}) - {\rm log}(\lambda_{k, el})$ corresponds to $d\lambda
\approx$ 0.017 ${\rm \AA}$, i.e., the wavelength pixel size on the
detector at the
center of the spectral coverage\footnote{The actual spectral
resolution at this point is $\approx$ 0.14 ${\rm \AA}$,
equivalent to just about 8 detector pixels.}; $N_{e} = 33$ is the number of
exposures, whereas $N_{l} = 123$ is the number of selected stellar
absorption lines. $Z_e$ stands for the ``redshift'' of the
$e^{th}$ exposure, and $S^{Z_e}$ denotes the spectrum that has
been Doppler-shift corrected according to the radial velocity of
the $e^{th}$ exposure. Wavelength solutions of different exposures
agree with each other to within $\sim$0.03 ${\rm \AA}$, therefore
the central wavelength of each stellar line, $\lambda_{0, el}$, is
a function of both line and exposure. We account for the small
variations in wavelength solutions by localizing each line in each
exposure separately.  In other words, $C^Z(k)$ is simply the
average scattered spectral line generated by summing, after
continuum normalization, over all of the selected lines (123 for
our HD 209458 data) and all of the exposures (33 for our HD 209458
data), a total of $N_{e}\times N_{l}$ separate spectral fragments
(4059 in the case of the present data).

Fourth, we take the mean (as well as the median) values calculated
from the same 33 pre-processed spectra, Doppler-shift corrected
using $N$ sets of scrambled redshifts (hereafter, reference
spectra, for short), as the standard to compare $C^Z(k)$ with.
This spectrum can be regarded as the ``continuum'' level, i.e.,
relatively free of scattered signals, and is given by
\begin{equation}
\langle C^{Z'}(k) \rangle = \frac{1}{N}\sum_{n = 1}^N C^{Z'_n}(k),
\end{equation}
where $C^{Z'_n}(k) = \frac{1}{N_{e}N_{l}}\sum_{e = 1}^{N_{e}}
\sum_{l = 1}^{N_{l}} S^{Z'_{n, e}}[{\rm log}(\lambda_{k, el}) -
{\rm log}(\lambda_{0, el})]$. $\{Z'_{n, e}\}$ is a permutation of
$\{Z_e\}$, and there are $N = 1,000$ sets of random permutations.
We note it is essential that $\{Z'_{n, e}\}$ is a permutation of
$\{Z_e\}$, rather than purely random redshifts, because using a
permutation of $\{Z_e\}$ ensures the exact same fragments of
stellar spectra go into $C^{Z'_n}(k)$, as those in $C^Z(k)$.

Fifth and finally, we define the combined scattered signal as the
residual given by
\begin{equation}
\Delta C(k) \equiv C^Z(k) - \langle C^{Z'}(k) \rangle.
\end{equation}
Given scattered light, $\Delta C(k)$ will be negative around $k =
0$, with a profile similar to the combined stellar lines scaled by
$\epsilon(\alpha, \lambda)$. Strictly speaking, $\langle C^{Z'}(k)
\rangle$ is affected to some degree by the scattered-light signal,
even though the redshifts have been scrambled. This is because
occasionally the randomly assigned redshift will happen to be or
close to the appropriate redshift of the exposure. Therefore
$\langle C^{Z'}(k) \rangle$ serves as a lower limit for the true
continuum level, and $\Delta C(k)$ yields a conservative value.

\subsubsection{Results for Strategy I}

\begin{figure*}
  \centering
    \includegraphics[width=120mm]{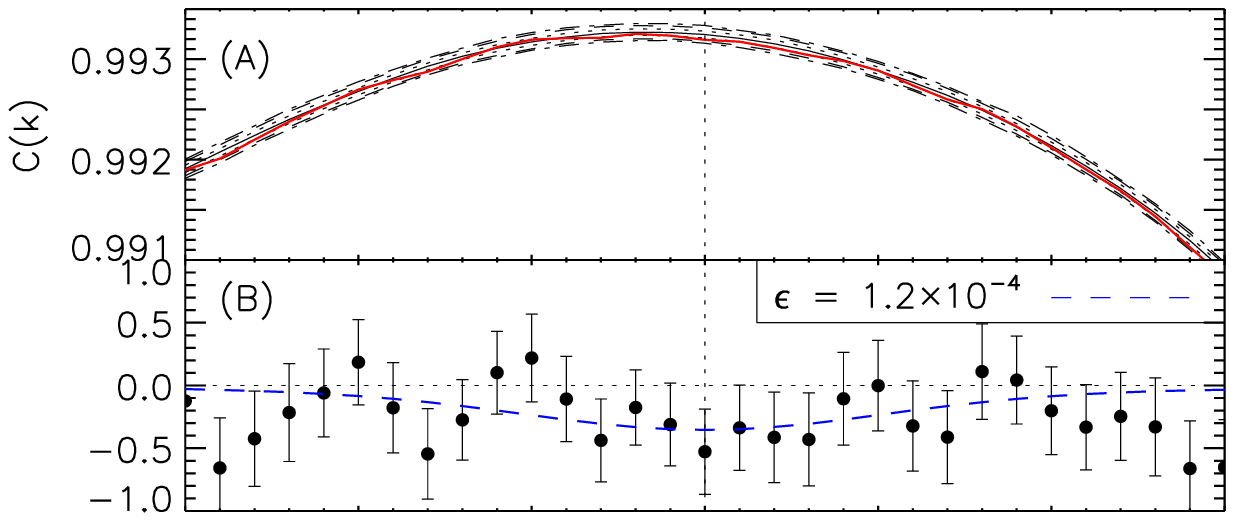}
    \includegraphics[width=120mm]{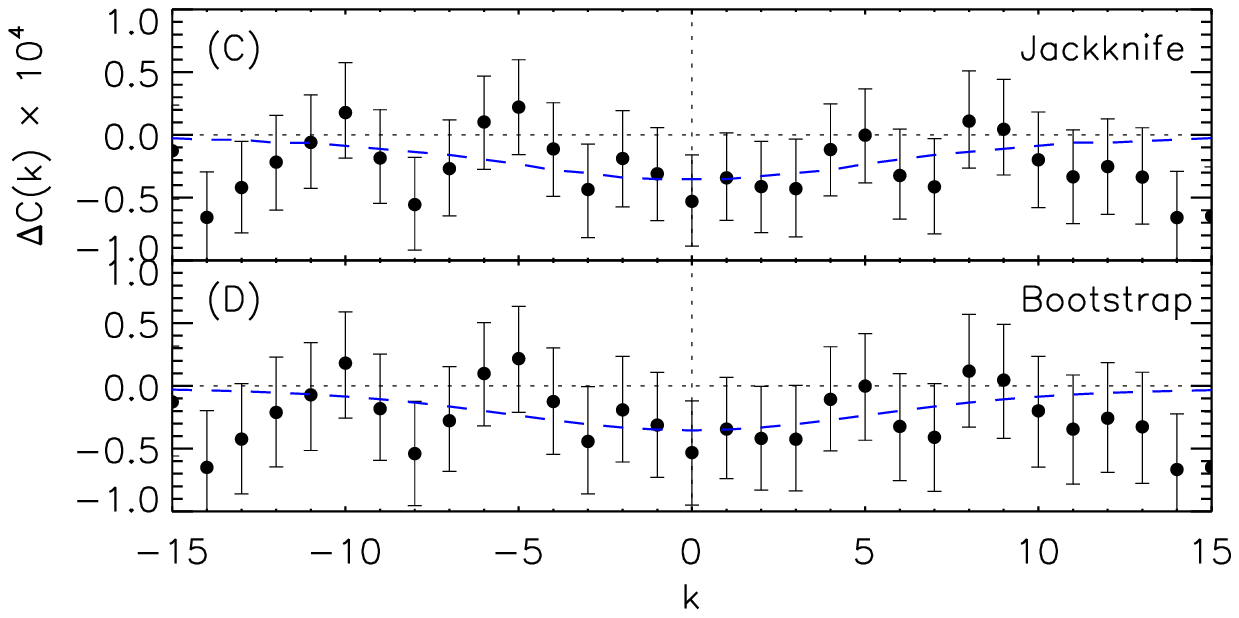}
    \caption{The normalized count $C(k)$ and the residual $\Delta C(k)$ as a
    function of wavelength difference $k$, relative to the predicted scattered line's
    center, aligned and average-combined over a large set of stellar lines and over
    exposures. Each exposure has been Doppler-shift corrected to align the scattered
    signals, according to the planet's relative radial velocity to the star.
    (A): the red curve shows $C^Z(k)$, whereas the black curves represent the
    median ({\it solid}; also same as the mean) and 68\% ({\it dotted}), 95\% ({\it dashed}),
    and 99\% ({\it dash-dotted}) significance levels, obtained from 1,000 sets of
    scrambling-shifted spectra ($\{C^{Z'_n}(k)\}$); (B), (C) and (D): the black
    filled-circles show the residual $\Delta C(k)$, which is the difference between
    $C^Z(k)$ ({\it red}) and $\langle C^{Z'}(k)\rangle$ ({\it black-solid}). Results
    from the simulated scattered light by combining the stellar lines scaled by a
    factor of $1.2\times10^{-4}$ are shown as {\it blue-dashed} curves. The error bars in (B)
    correspond to 1 $\sigma$- errors obtained from the significance levels shown in (A),
    whereas (C) and (D) present the results of 4,059 sets of {\sf jackknife}- and {\sf bootstrap}-
    re-samplings. In all the panels, one pixel corresponds to $\sim 0.017 {\rm \AA}$, the approximate size of a detector pixel, about one eighth of 
the actual spectral resolution.  The systematic decrease in $\Delta C(k)$ at more than 12 pixels to the red or blue of the line center is due to the
influence of stellar absorption lines beyond the edges of the region
chosen to be clear of such interference.}
    \label{fig:subfig:nc}
\end{figure*}

\begin{figure*}
  \centering
    \includegraphics[width=120mm]{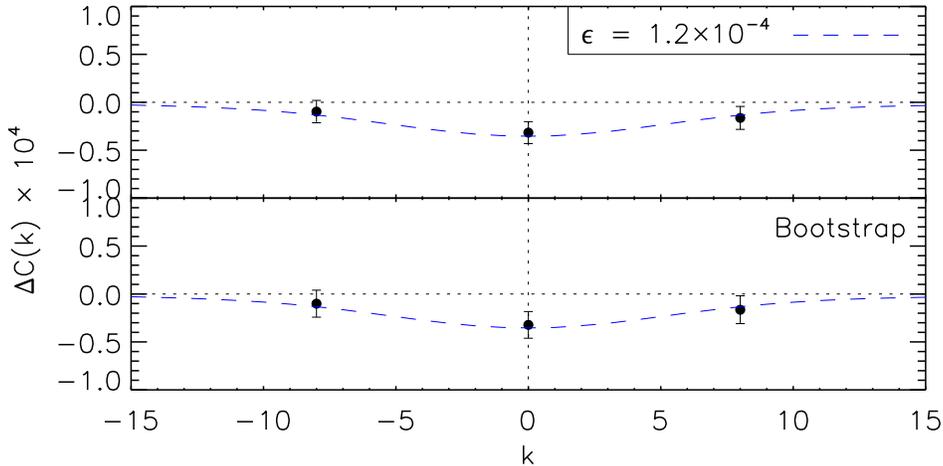}
    \caption{The normalized residual counts $\Delta C(k)$ as a function of 
wavelength difference $k$, as in panels (B) and (D) of the previous figure,
but in this case the data points have been rebinned by a factor of 8.
This corresponds to the true spectral resolution of the data, rather than
to the detector pixel size.  The rebinnig is arranged such that the 
predicted line center position at $k = 0$ is in the middle of the central
point.  The dashed line shows the expected shape of the composite scattered
light line assuming a $1.2\times10^{-4}$ scattered light fraction. }
    \label{fig:subfig:ncrebin}
\end{figure*}

Figure \ref{fig:subfig:nc} shows the normalized count $C(k)$ and
the residual $\Delta C(k)$ obtained from analyzing the Subaru/HDS
spectra of HD 209458b. The overall shape of $C(k)$ is a result of
the combination of multiple stellar spectrum segments. The
significance levels plotted in (A), as well as the 1-$\sigma$
errors shown in (B) are calculated from the variance among the
reference spectra. $\Delta C(k)$ along with the 1-$\sigma$ errors
in (C) and (D) are from the jackknife- and bootstrap- re-samplings
(re-sampled over both lines and exposures) of the data along with
the corresponding reference spectra, respectively. The re-sampling
tests produced similar errors to those solely estimated from the
variations in reference spectra. To test the algorithms, we also
analyzed a mock/simulated spectrum, obtained by combining the same
set of stellar lines scaled by a factor of $1.2\times10^{-4}$.
This simulated signal indicates the expected scattered light
assuming a planet-to-star flux ratio of $\epsilon =
1.2\times10^{-4}$. The comparison of the observations and
simulated signal shown in Figure \ref{fig:subfig:nc} is intriguing
in that the data are consistent with a low statistical
significance detection of the scattered light, with $\epsilon
\gtrsim 10^{-4}$ at the 1-$\sigma$ level, as the data follow the
simulated signal tantalizingly well, especially around the
predicted line center where the signal should be the strongest.

Figure \ref{fig:subfig:ncrebin} also displays the normalized residual
counts $\Delta C(k)$, but in this figure the actual spectral resolution,
about 8 detector pixels, is used.  There are thus only three points
across the expected composite scattered light absorption line, but the
error bars are correspondingly reduced, and the points are not strongly
correlated by the finite spectral resolution as they are in Figure
\ref{fig:subfig:nc}. Since the two figures simply display the same data at
two different wavelength resolutions, they are both naturally
consistent with the same scattered-light fraction of $1.2\times10^{-4}$
and the same line shape, shown by the {\it blue-dashed} line.

\subsection{Strategy II: Equivalent Width Ratio Distributions}\label{sec:strat2}

In this section, we further suggest a method to measure the
planet-to-star flux ratio $\epsilon$ by examining the distribution
of the EW ratios between the scattered and the stellar absorption
lines, measured for a large set of lines and exposures. The FWHMs
of stellar lines displayed in Figure \ref{fig:fwhm} are used to
set the wavelength range over which the EWs of both stellar and
scattered lines are evaluated.  Since this method effectively
concentrates all of the ``missing flux'' signal into a single
number with appropriate weighting of the stronger and weaker
features, we would expect better sensitivity performance than
Strategy I.  The cost, of course, is loss of all information about
the shape of the composite shape of the scattered lines and thus
of a useful check that the measured feature has the expected
character of scattered light.  The two approaches are thus
complementary.

\subsubsection{Details of the Strategy-II Procedure}

We adopt a dimensionless EW, which is convenient for a
logarithmically linear wavelength scale, given by
\begin{equation}
W \equiv \int
\frac{d\lambda}{\lambda_0}\bigg[1-\frac{S_{\lambda}}{S(0)}\bigg] =
{\rm ln}10 \cdot \int \bigg[1-\frac{S_{{\rm
log}\lambda}}{S(0)}\bigg]d{\rm log}\lambda,
\end{equation}
where $\lambda_0$ is the central wavelength of a stellar
absorption line and $S(0)$ denotes its local continuum. In
logarithmically linear wavelength scale, $W$ can be discretized as
\begin{equation}
W = A \cdot \sum_i \bigg[1 - \frac{S_{{\rm log}\lambda_i}}{S(0)}\bigg],
\end{equation}
where $A\equiv{\rm ln}10\cdot \Delta{\rm log}\lambda$ is a
constant, and the summation goes over all pixels inside a
wavelength range of twice that of the line's FWHM and centered at
$\lambda_0$. Since different stellar lines have different FWHMs as
shown in Figure \ref{fig:fwhm}, a variable number of pixels are
included in the EW integrations by making use of the line-strength
information, whereas in Strategy I, the same number of pixels are
combined for all lines.

The EW of the stellar lines, $W^{\star}_{e, l}=A\cdot\sum_i [1 -
S^0_{{\rm log}\lambda_{i, el}}/S^0_{e, l}(0)]$, is calculated for
each line (denoted by $l$) in each exposure (denoted by $e$),
using the stellar FWHMs measured in the initial exposure, which
are shown in Figure \ref{fig:fwhm}. To check for FWHM variability,
we measure the FWHMs in both the initial and final exposures and
find that they agreed within 0.1 per cent.

For each line, the local continuum $S^0_{e, l}(0)$ is estimated by
averaging the median of the counts in two spectral intervals on
either side of the line, with a size chosen to be neither too
large to reflect the local continuum level nor too small for
accurate statistics. We use a range of 50 pixels, corresponding to
$\sim$1 ${\rm \AA}$. The results have been tested to be
independent of the adopted size of the range over which the
continuum levels are estimated.

For the scattered signals, $W^{Z_e}_{e, l}=A\cdot\sum_i [1 -
S^{Z_e}_{{\rm log}\lambda_{i, el}}/S^{Z_e}_{e,l}(0)]$, where the
super script $Z_e$ represents the spectrum $S$ has been
Doppler-shift corrected with the Doppler shift of the $e^{th}$
exposure. We use the FWHMs of the stellar lines, assuming that the
scattered light is a replica of starlight, scaled by a factor of
$\epsilon$. The local continuum $S^{Z_e}_{e,l}(0)$ is taken as the
median of the line's blue end, since the red end often overlaps
the associated strong stellar lines. The EW ratio is then given by
$R^{Z_e}_{e, l} \equiv W^{Z_e}_{e, l} / W^{\star}_{e, l}$. The
EW-ratio measurements of $N_l$ lines in $N_e$ exposures constitute
a statistical sample of $\{R^{Z_e}_{e, l}\}$, which has $N_l\times
N_e$ data points.

We also calculate $\{R^{Z'_n}_{e, l}\}$ for $N$ sets of reference
spectra, to construct a standard comparison pool. Both {\sf mean}
and {\sf median} are adopted as statistical measures, where
$M(R^{Z})$ denotes taking either the mean or the median of
$\{R^{Z_e}_{e, l}\}$. Finally, we define the residual EW ratio,
given by
\begin{eqnarray}
\Delta M(R) \equiv M(R^{Z}) - \langle M(R^{Z'})\rangle, \nonumber \\
         {\rm where} \,\,\, \langle M(R^{Z'})\rangle = \frac{1}{N}\sum_{n = 1}^N M(R^{Z'_n}),
\end{eqnarray}
is the comparison standard obtained from $N$ sets of
scrambling-shifted reference spectra. Given null (by construction)
reference-spectrum signal, $\Delta M(R)$ will be positive and give
a direct estimate of $\epsilon$.

\subsubsection{Results for Strategy II}

%
\begin{figure*}
  \centering
    \includegraphics[width=150mm]{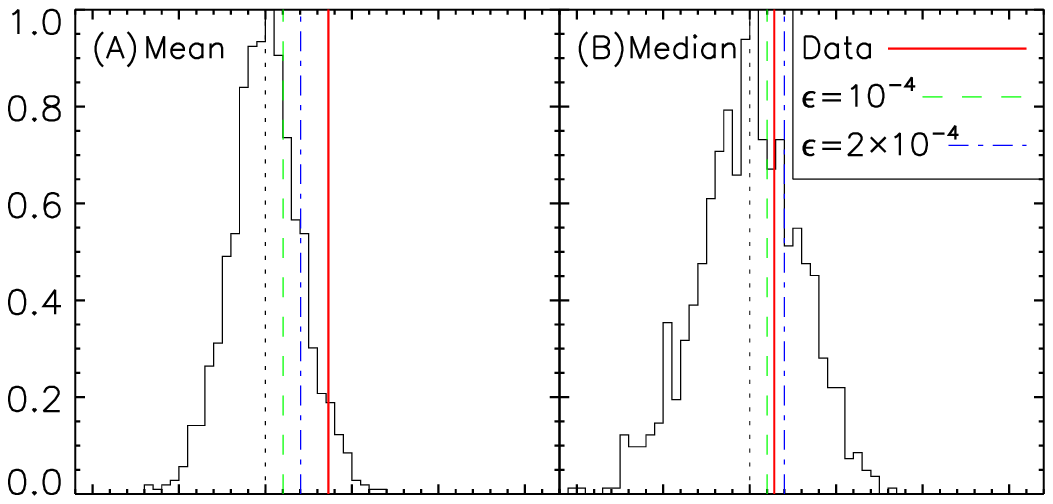}
    \includegraphics[width=150mm]{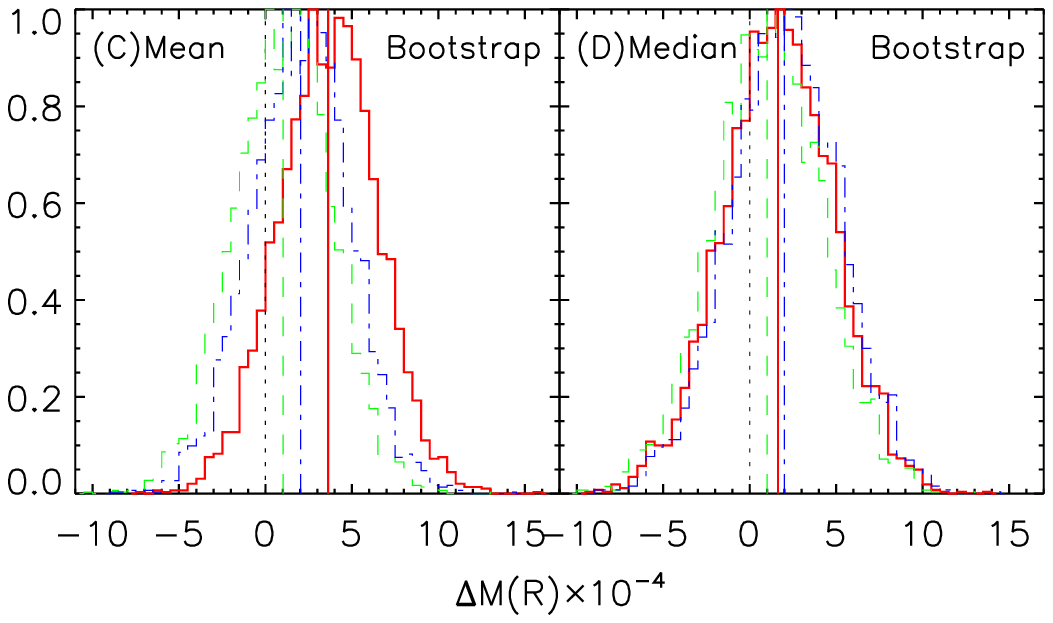}
    \caption{The EW ratio $\Delta M(R)$. (A) and (C) show the {\sf mean} results whereas
    (B) and (D) present the {\sf median}. In (A) and (B), the histogram displays the normalized
    distribution of results from 1,000 sets of scrambling-shifted spectra. The
    {\it dotted} line represents the averages of the distributions.
    As shown by the legend in (B), the other three vertical lines show results calculated
    from the data ({\it thick red-solid}), the simulated scattered-light spectra assuming
    $\epsilon = 10^{-4}$ ({\it thin green-dashed}), and $\epsilon = 2\times10^{-4}$
    ({\it thin blue-dash-dotted}), respectively. (C) and (D) show the normalized distributions
    of results from 4,059 {\sf bootstrap}- re-samplings, for the data ({\it thick red-solid}), the
    simulated scattered-light spectra assuming $\epsilon = 10^{-4}$ ({\it thin green-dashed}),
    and $\epsilon = 2\times10^{-4}$ ({\it thin blue-dash-dotted}), respectively. The three
    vertical lines mark the averages of the normalized distributions.
    $\Delta M(R) = 0$ is also shown as {\it dotted} lines for comparison.}
    \label{fig:subfig:ew}
\end{figure*}
%

Figure \ref{fig:subfig:ew} shows $\Delta M(R)$ for both the median
and the mean statistics, obtained from the HD 209458b data,
simulated scattered signals, reference spectra, and bootstrap
re-samplings. We have also generated jackknife re-samplings, which
produced similar results to those of the bootstrap re-samplings,
and both re-sampling analysis yield consistent errors with those
estimated from the variation of reference spectra. To test the
linearity of the algorithm, we injected mock/simulated scattered
signals, assuming $\epsilon = 10^{-4}$ and $2\times10^{-4}$,
respectively. The injected signals are recovered linearly through
the data pipeline as expected, since our algorithm only involves
linear transformations. These EW-ratio results suggest $\epsilon =
(1.4\pm2.9)\times10^{-4}$ ($(3.6\pm2.0)\times10^{-4}$), according
to the median (mean) statistics. The mean is more sensitive to
outliers which can be caused by stellar-line overlaps or
inaccurate estimation of the local continuum levels, or by real
intrinsic variation of the albedo with wavelength or phase,
whereas the median should be more robust and representative of a
global effective albedo over the whole wavelength and phase
ranges.

\section{Summary and Discussion}

We have developed a simple, robust, linear, and model-independent
technique to detect scattered starlight from transiting CEGPs, by
measuring scattered spectral features at predictable Doppler
shifts from the stellar lines due to planets' orbital motions.
There are two main variations of this technique. First, spectrum
fragments have been aligned and combined over a large set of
stellar absorption lines as well as over many exposures taken near
the secondary eclipse. Second, the planet-to-star flux ratio has
been determined by examining the distribution of the EW ratios
between the scattered and stellar absorption lines. We have made a
comparison sample by Doppler-shift correcting the exposures with
multiple sets of scrambled, instead of the real, redshifts. This
comparison sample has been passed through the same data pipeline,
and provided the control comparisons, as well as the significance
levels of any detections. Jackknife- and bootstrap- re-samplings
have yielded consistent errors with those determined from the
variations of the reference spectra. We have also produced spectra
with simulated scattered signals, using the stellar absorption
lines scaled by a factor of an assumed planet-to-star flux ratio.
The simulated signals have been fully recovered through the data
pipeline.  Physical parameters of the system are required to have
been determined, in order to predict the planet's relative radial
velocity at each orbital phase, and further localize its imprinted
signals on the recorded spectra. These parameters can be
determined from radial-velocity and photometric observations of
transiting CEGPs with high precision and reliability.

As a case study, we have analyzed very high S/N optical spectra of
HD 209458b, acquired with the Subaru HDS in UT 2002 October. Our
results suggest an average planet-to-star flux ratio of
(1.4$\pm$2.9)$\times10^{-4}$, in the wavelength from 554 nm to 681
nm, when the planet was during the orbital phases $11.0^{\circ} <
\alpha < 33.9^{\circ}$. Assuming a Lambert-sphere phase function
\citep{pollack86,charbonneau99}, the equivalent geometric albedo
can be estimated as $0.8\pm1.6$, according to Equation
\ref{equ:ratio}, where we take the phase of the middle exposure
$\phi \approx 0.561$, corresponding to $\alpha = 22.3^{\circ}$,
out of $11.0^{\circ} < \alpha < 33.9^{\circ}$ for all the
exposures we used. The best previous existing constraint on
scattered light from HD 209458b is based on direct imaging with
the $MOST$ satellite \citep{rowe06}. \citet{rowe06} found a
1-$\sigma$ limit on the planet-to-star flux ratio of
$4.88\times10^{-5}$, corresponding to a geometric albedo upper
limit in $MOST$ bandpass (400-700 nm) of 0.25. Our results provide
a useful point of comparison for this difficult measurement,
covering a  somewhat different wavelength range and based on
different observational strategies and analysis techniques.

These results do not seem yet able to support a comparison to
models of the HD 209458b atmosphere or to justify detailed
physical interpretation, but they do not miss the mark by far and
are thus quite encouraging.  Specifically, the Subaru HDS data
analyzed here constitute well under 5 hours of total exposure
time; it would be possible to obtain more data, and there is no
indication in the present data of systematic problems (such as
instrumental instabilities) that would prevent further integration
from producing a greater sensitivity. Moreover, one significantly
more favorable transiting CEGP system (HD 189733) is known, and
others may be discovered. In combination with measurements of
re-emitted thermal radiation \citep[e.g.][]{charbonneau05},
scattered light studies offer the possibility of major qualitative
advances in our understanding of CEGPs, a central riddle in the
study of exoplanets generally.

\acknowledgments This work is based on data from the Subaru
Telescope, which is operated by the National Astronomical
Observatory of Japan. We wish to recognize and acknowledge the
very significant cultural role and reverence that the summit of
Mauna Kea has always had within the indigenous Hawaiian community.
We are most fortunate to have the opportunity to conduct
observations from this mountain.

\bibliographystyle{apj}
\bibliography{apj-jour,pslrefs}

\begin{thebibliography}{24}
\expandafter\ifx\csname natexlab\endcsname\relax\def\natexlab#1{#1}\fi

\bibitem[{{Brown} {et~al.}(2001){Brown}, {Charbonneau}, {Gilliland}, {Noyes},
  \& {Burrows}}]{brown01}
{Brown}, T.~M., {Charbonneau}, D., {Gilliland}, R.~L., {Noyes}, R.~W., \&
  {Burrows}, A. 2001, \apj, 552, 699

\bibitem[{{Burrows} {et~al.}(2005){Burrows}, {Hubeny}, \&
  {Sudarsky}}]{burrows05}
{Burrows}, A., {Hubeny}, I., \& {Sudarsky}, D. 2005, \apjl, 625, L135

\bibitem[{{Cameron} {et~al.}(1999){Cameron}, {Horne}, {Penny}, \&
  {James}}]{cameron99}
{Cameron}, A.~C., {Horne}, K., {Penny}, A., \& {James}, D. 1999, \nat, 402, 751

\bibitem[{{Charbonneau} {et~al.}(2005){Charbonneau}, {Allen}, {Megeath},
  {Torres}, {Alonso}, {Brown}, {Gilliland}, {Latham}, {Mandushev}, {O'Donovan},
  \& {Sozzetti}}]{charbonneau05}
{Charbonneau}, D., {Allen}, L.~E., {Megeath}, S.~T., {Torres}, G., {Alonso},
  R., {Brown}, T.~M., {Gilliland}, R.~L., {Latham}, D.~W., {Mandushev}, G.,
  {O'Donovan}, F.~T., \& {Sozzetti}, A. 2005, \apj, 626, 523

\bibitem[{{Charbonneau} {et~al.}(2000){Charbonneau}, {Brown}, {Latham}, \&
  {Mayor}}]{charbonneau00}
{Charbonneau}, D., {Brown}, T.~M., {Latham}, D.~W., \& {Mayor}, M. 2000, \apjl,
  529, L45

\bibitem[{{Charbonneau} {et~al.}(1999){Charbonneau}, {Noyes}, {Korzennik},
  {Nisenson}, {Jha}, {Vogt}, \& {Kibrick}}]{charbonneau99}
{Charbonneau}, D., {Noyes}, R.~W., {Korzennik}, S.~G., {Nisenson}, P., {Jha},
  S., {Vogt}, S.~S., \& {Kibrick}, R.~I. 1999, \apjl, 522, L145

\bibitem[{{Chauvin}(2007)}]{chauvin07}
{Chauvin}, G. 2007, in Proceedings of the conference In the Spirit of Bernard
  Lyot: The Direct Detection of Planets and Circumstellar Disks in the 21st
  Century. June 04 - 08, 2007. University of California, Berkeley, CA, USA.
  Edited by Paul Kalas., ed. P.~{Kalas}

\bibitem[{{Collier Cameron} {et~al.}(2002){Collier Cameron}, {Horne}, {Penny},
  \& {Leigh}}]{cameron02}
{Collier Cameron}, A., {Horne}, K., {Penny}, A., \& {Leigh}, C. 2002, \mnras,
  330, 187

\bibitem[{{Henry} {et~al.}(2000){Henry}, {Marcy}, {Butler}, \&
  {Vogt}}]{henry00}
{Henry}, G.~W., {Marcy}, G.~W., {Butler}, R.~P., \& {Vogt}, S.~S. 2000, \apjl,
  529, L41

\bibitem[{{Knutson} {et~al.}(2007){Knutson}, {Charbonneau}, {Noyes}, {Brown},
  \& {Gilliland}}]{knutson07}
{Knutson}, H.~A., {Charbonneau}, D., {Noyes}, R.~W., {Brown}, T.~M., \&
  {Gilliland}, R.~L. 2007, \apj, 655, 564

\bibitem[{{Leigh} {et~al.}(2003){Leigh}, {Cameron}, \& {Guillot}}]{leigh03}
{Leigh}, C., {Cameron}, A.~C., \& {Guillot}, T. 2003, \mnras, 346, 890

\bibitem[{{Marley} {et~al.}(1999){Marley}, {Gelino}, {Stephens}, {Lunine}, \&
  {Freedman}}]{marley99}
{Marley}, M.~S., {Gelino}, C., {Stephens}, D., {Lunine}, J.~I., \& {Freedman},
  R. 1999, \apj, 513, 879

\bibitem[{{Mayor} \& {Queloz}(1995)}]{mayor95}
{Mayor}, M. \& {Queloz}, D. 1995, \nat, 378, 355

\bibitem[{{Mazeh} {et~al.}(2000){Mazeh}, {Naef}, {Torres}, {Latham}, {Mayor},
  {Beuzit}, {Brown}, {Buchhave}, {Burnet}, {Carney}, {Charbonneau}, {Drukier},
  {Laird}, {Pepe}, {Perrier}, {Queloz}, {Santos}, {Sivan}, {Udry}, \&
  {Zucker}}]{mazeh00}
{Mazeh}, T., {Naef}, D., {Torres}, G., {Latham}, D.~W., {Mayor}, M., {Beuzit},
  J.-L., {Brown}, T.~M., {Buchhave}, L., {Burnet}, M., {Carney}, B.~W.,
  {Charbonneau}, D., {Drukier}, G.~A., {Laird}, J.~B., {Pepe}, F., {Perrier},
  C., {Queloz}, D., {Santos}, N.~C., {Sivan}, J.-P., {Udry}, S., \& {Zucker},
  S. 2000, \apjl, 532, L55

\bibitem[{{Narita} {et~al.}(2005){Narita}, {Suto}, {Winn}, {Turner}, {Aoki},
  {Leigh}, {Sato}, {Tamura}, \& {Yamada}}]{narita05}
{Narita}, N., {Suto}, Y., {Winn}, J.~N., {Turner}, E.~L., {Aoki}, W., {Leigh},
  C.~J., {Sato}, B., {Tamura}, M., \& {Yamada}, T. 2005, \pasj, 57, 471

\bibitem[{{Noguchi} {et~al.}(2002){Noguchi}, {Aoki}, {Kawanomoto}, {Ando},
  {Honda}, {Izumiura}, {Kambe}, {Okita}, {Sadakane}, {Sato}, {Tajitsu},
  {Takada-Hidai}, {Tanaka}, {Watanabe}, \& {Yoshida}}]{noguchi02}
{Noguchi}, K., {Aoki}, W., {Kawanomoto}, S., {Ando}, H., {Honda}, S.,
  {Izumiura}, H., {Kambe}, E., {Okita}, K., {Sadakane}, K., {Sato}, B.,
  {Tajitsu}, A., {Takada-Hidai}, T., {Tanaka}, W., {Watanabe}, E., \&
  {Yoshida}, M. 2002, \pasj, 54, 855

\bibitem[{{Perryman}(2000)}]{perryman00}
{Perryman}, M.~A.~C. 2000, Reports of Progress in Physics, 63, 1209

\bibitem[{{Pollack} {et~al.}(1986){Pollack}, {Rages}, {Baines}, {Bergstralh},
  {Wenkert}, \& {Danielson}}]{pollack86}
{Pollack}, J.~B., {Rages}, K., {Baines}, K.~H., {Bergstralh}, J.~T., {Wenkert},
  D., \& {Danielson}, G.~E. 1986, Icarus, 65, 442

\bibitem[{{Rowe} {et~al.}(2006){Rowe}, {Matthews}, {Seager}, {Kuschnig},
  {Guenther}, {Moffat}, {Rucinski}, {Sasselov}, {Walker}, \& {Weiss}}]{rowe06}
{Rowe}, J.~F., {Matthews}, J.~M., {Seager}, S., {Kuschnig}, R., {Guenther},
  D.~B., {Moffat}, A.~F.~J., {Rucinski}, S.~M., {Sasselov}, D., {Walker},
  G.~A.~H., \& {Weiss}, W.~W. 2006, \apj, 646, 1241

\bibitem[{{Seager} \& {Sasselov}(1998)}]{seager98}
{Seager}, S. \& {Sasselov}, D.~D. 1998, \apjl, 502, L157+

\bibitem[{{Seager} {et~al.}(2000){Seager}, {Whitney}, \& {Sasselov}}]{seager00}
{Seager}, S., {Whitney}, B.~A., \& {Sasselov}, D.~D. 2000, \apj, 540, 504

\bibitem[{{Sudarsky} {et~al.}(2003){Sudarsky}, {Burrows}, \&
  {Hubeny}}]{sudarsky03}
{Sudarsky}, D., {Burrows}, A., \& {Hubeny}, I. 2003, \apj, 588, 1121

\bibitem[{{Sudarsky} {et~al.}(2000){Sudarsky}, {Burrows}, \&
  {Pinto}}]{sudarsky00}
{Sudarsky}, D., {Burrows}, A., \& {Pinto}, P. 2000, \apj, 538, 885

\bibitem[{{Winn} {et~al.}(2004){Winn}, {Suto}, {Turner}, {Narita}, {Frye},
  {Aoki}, {Sato}, \& {Yamada}}]{winn04}
{Winn}, J.~N., {Suto}, Y., {Turner}, E.~L., {Narita}, N., {Frye}, B.~L.,
  {Aoki}, W., {Sato}, B., \& {Yamada}, T. 2004, \pasj, 56, 655

\end{thebibliography}

\end{document}